\documentclass[twocolumn,english,prl,showpacs,aps]{revtex4-1}
\usepackage[T1]{fontenc}
\usepackage[latin9]{inputenc}
\usepackage{amssymb}
\usepackage{times}
\usepackage{graphicx}
\usepackage{amsmath}
\usepackage{pstricks}
\usepackage{color}
\usepackage{subfigure}

\begin{document}
\title{Imaging single Rydberg electrons in a Bose-Einstein condensate}

\author{Tomasz Karpiuk,$^{1,2}$ Miros{\l}aw Brewczyk,$^{1,3}$ and Kazimierz Rz\k a\.zewski$^{3,4}$}
\affiliation{\mbox{$^1$Wydzia{\l} Fizyki, Uniwersytet w Bia{\l}ymstoku,
                       ul. Lipowa 41, 15-424 Bia{\l}ystok, Poland}  \\
\mbox{$^2$Centre for Quantum Technologies, National University of Singapore, 3 Science Drive 2, Singapore 117543, Singapore }  \\
\mbox{$^3$Center for Theoretical Physics PAN, Al. Lotnik\'ow 32/46, 02-668 Warsaw, Poland}  }
%%%%%%%%%%%%%%%%%%%%%%%%%%%
\author{Anita Gaj$^{4}$, Jonathan B. Balewski$^{4}$, Alexander T. Krupp$^{4}$, Michael~Schlagm\"{u}ller$^{4}$, Robert L\"{o}w$^{4}$, Sebastian Hofferberth$^{4}$ and Tilman Pfau$^{4}$}
\affiliation{
\mbox{$^4$5. Physikalisches Institut and Center for Integrated Quantum Science and Technology IQST,}\\ \mbox{Universit\"at Stuttgart, Pfaffenwaldring 57, 70569 Stuttgart, Germany} }
%%%%%%%%%%%%%%%%%%%%%%%%%%%%
\begin{abstract}
The quantum mechanical states of electrons in atoms and molecules are distinct 
orbitals, which are fundamental for our understanding of atoms, molecules and 
solids. Electronic orbitals determine a wide range of basic atomic properties, 
allowing also for the explanation of many chemical processes. Here, we propose a 
novel technique to optically image the shape of electron orbitals of neutral atoms 
using electron-phonon coupling in a Bose-Einstein condensate. To validate our model 
we carefully analyze the impact of a single Rydberg electron onto a condensate and 
compare the results to  experimental data. Our scheme requires only well-established 
experimental techniques that are readily available and allows for the  direct 
capture of textbook-like spatial images of single electronic orbitals in a single 
shot experiment.
\end{abstract}
%%%%%%%%%%%%%%%%%%%%%%%%%%%%%%%%
%\pacs{???}

%\date{\today}
\maketitle
The wavefunction is a fundamental concept of quantum mechanics.
Our current understanding of atoms, molecules and solids is 
based on the fact that the probability density of an electron is the absolute 
square of the electronic wavefunction. The theoretical description of electron 
orbitals was founded at the beginning of the last century. Yet, to date the 
spatial structure of most orbitals has not been observed directly. Most current 
techniques to study the wavefunction of electrons in atoms and molecules rely on 
tomographic reconstruction. Based on this technique the wavefunction of 
the energetically highest orbital of various molecules has been obtained using 
high harmonics generated in the interaction of intense femtosecond laser pulses
with molecules~\cite{ILZ04} and photoemission spectroscopy~\cite{PBF09}. Another method, 
which has been until now applied to larger polymers, is based on scanning tunneling 
microscopy~\cite{RMS05, GMM11, CDG11}. Furthermore, an image of the electron 
wavefunction in single hydrogen atoms has been reconstructed only recently~\cite{SRL13}. 
There, electronic states in a strong static electric field have been observed via 
photoionization and subsequent electron detection using a magnifying electrostatic 
lens. \\
%Direct spatial imaging of single atomic orbitals, as they are known from basic quantum theory, 
%has, however, not been realized yet. \\
\indent
Here, we propose a method to optically image the orbitals of electrons excited 
to a Rydberg state. These orbitals are larger in size than optical wavelengths.
Moreover, due to the different sizable quantum defects and dipolar selection rules, they 
can be prepared in a well-defined quantum state providing clean s, p, d and f series. 
The proposed method is based on the interaction of the Rydberg 
electron with a dense ultracold gas~\cite{LossesNature}. Due to this interaction the probability 
density of a single Rydberg electron can be imprinted on the density of
surrounding Bose-Einstein condensate (BEC) atoms. Thus, textbook-like optical images of hydrogenic 
states can be obtained using already well-established 
imaging techniques for cold atoms. \\
\indent
In order to model the Rydberg excitation 
dynamics and the phase imprint onto a finite-size BEC we develop a numerical model 
describing the probabilistic Rydberg excitation process and the subsequent interaction 
with the finite-size BEC. Our approach agrees well with available experimental 
data on Rydberg excitations in a BEC \cite{LossesNature} and confirms electron-phonon 
coupling as the underlying mechanism, which has been studied previously in the 
framework of Bogoliubov approximation. We discuss the experimental requirements and 
challenges to implement our proposal, including finite imaging resolution as well 
as the role of atomic and photonic shot noise for the expected images of Rydberg 
orbitals.\\
\indent 
The $1/r^4$ interaction between the single electron of the Rydberg atom and 
polarizable ground state atoms can be well described by a pseudopotential~\cite{Fermi, F36} 
resulting in an effective potential acting on the ground state atoms of the form
\begin{equation}
 V_{Ryd}(\vec{r}) = \frac{2\pi \hbar^2 a}{m_e} |\Psi_{Ryd}(\vec{r})|^2 \,,
\label{Ryd}
\end{equation}	
where $\Psi_{Ryd}(\vec{r})$ is the Rydberg electron wavefunction, $a$ denotes the 
electron-atom $s$-wave triplet scattering length, $a = -16.1$\,a.\,u. for 
$^{87}$Rb~\cite{Bahrim01}, and $m_e$ is the electron mass. This well-known model 
has previously been used to make quantitative statements about the binding energy 
and excitation spectra of ultralong-range Rydberg molecules~\cite{Vera, Alex, Shaffer, Raithel}. 
Higher partial waves are irrelevant for principal quantum number $n>100$. 
The interaction between the ionic core and the BEC is \textasciitilde300 times 
smaller and can be safely neglected~\cite{Pethick}. Thus, the interaction between 
the Rydberg atom and ground state atoms creates a potential $V_{Ryd}$ around the 
excited atom with a structure defined by the Rydberg electron orbital. \\ 
\indent 
To model the effect of a single Rydberg electron on the BEC we introduce the 
pseudopotential term $V_{Ryd}$ as a mean field component in the Gross-Pitaevskii 
equation (GPE). GPE describes the dynamics of the bosonic atomic field. We adopt 
a classical field approximation (CFA), where a long-wavelength atomic field is 
replaced by a classical complex function $\Psi(\vec{r},t)$ satisfying the time-dependent GPE
\begin{eqnarray}
 i\hbar \frac{\partial}{\partial t} \Psi(\vec{r},t) &=& \left[
-\frac{\hbar^2}{2m}\nabla^2
+ V_{trap}(\vec{r})
+ g|\Psi(\vec{r},t)|^2
\right. \nonumber  \\
&&\left.
+ f(t)\;V_{Ryd}(\vec{r}-\vec{R})
\right] \Psi(\vec{r},t)  \,,
\label{GPeq}
\end{eqnarray}
where $f(t)$ is 1 for the finite time when the Rydberg atom is present in the BEC 
and 0 otherwise. On the right-hand side the first three terms are related to the 
kinetic energy, the trapping potential and the contact interaction with coupling 
constant $g$. CFA is a valid treatment for describing Bogoliubov-Popov excitations \cite{Brewczyk}. \\
\indent
Before we turn to the investigation of the imaging of electron orbitals, we use our 
approach to model our recent experiment, where about a few hundred Rydberg atoms were excited 
successively at random positions inside a BEC. In the experiment \cite{LossesNature} 
a Rydberg atom in an $s$-state with principal quantum numbers~$n$ ranging 
from $110$ to $202$ was created in a condensate of $^{87}$Rb atoms. We used a 
1$\mu$s light pulse, during which the Rydberg atom got excited with a certain 
probability. 10$\,\mu$s after the excitation pulse, we sent a 2$\,\mu$s ionization  
pulse, which extracted the Rydberg atom unless it has not been lost before. Although 
the Rydberg blockade mechanism~\cite{Saffman} ensures that at any moment there was 
not more than a single excitation within the BEC, we studied only the cumulative effect of 
many successive excitations on the BEC. 
In the finite-size BEC the resonance frequency is modified by the spatially varying 
energy shift $\delta E (\vec{R},t)$~\cite{Anita} due to local density. 
We model this complicated many-body excitation process by a stochastic model. 
\\
\indent
Each appearance and disappearance of the potential $V_{Ryd}$ are examples of 
quantum jumps. We check if a randomly chosen atom on a grid representing the 
density distribution $\rho(\vec{r})$ is in a Rydberg state according to the excitation 
probability. The probability to find an atom at position~$\vec{R}$ in the Rydberg 
state, for sufficiently short times~$t$ and low single atom Rabi frequencies~$\Omega_R$, 
is given as
\begin{equation}
 p(\vec{R},t) = \frac{\Omega_R^2}{\Omega^2(\vec{R},t)}\sin^2{\left[\Omega(\vec{R},t) \, t /2 \right]},
\label{probability}
\end{equation}
where $\Omega(\vec{R},t)=\sqrt{\Omega_R^2+\Delta^2(\vec{R},t)}$ is the effective 
Rabi frequency, which accounts for a non-zero local detuning $\Delta(\vec{R},t)$. 
This spatially varying detuning~$\Delta(\vec{R},t)$ is given by a frequency 
difference of the detuning $\Delta \omega_{L}$ of the excitation laser from the 
Rydberg transition in an unperturbed atom and an additional mean field shift 
$\delta E (\vec{R},t)$. 
Since the condensate density changes due to the appearance of successive Rydberg 
atoms, the detuning~$\Delta$ and thus the excitation probability~(\ref{probability}) 
depend also on time. 
\\
\indent Following a coherent evolution at all possible grid points a localized 
Rydberg atom is potentially generated in our simulation on the time scale of the 
decoherence rate due to elastic scattering with a ground state atom. Therefore we 
choose the time step for our coherent evolution to be 200\,ns, which we then 
interrogate for the presence of a localized Rydberg atom. We repeat checking the 
atoms every 200\,ns until the end of the excitation pulse if no Rydberg atom 
was found in the previous iteration. Once a Rydberg atom is created we 
propagate the Gross-Pitaevskii equation with the $V_{Ryd}$ term included to calculate the 
evolution of the perturbed BEC while the Rydberg atom is present in the BEC. 
The interaction of a single Rydberg electron with surounding BEC atoms decays 
exponentially in time with a time constant of \textasciitilde10\,$\mu$s 
(see supplementary material, \cite{SM}).   After $13$\,$\mu$s 
the whole procedure starts again, however, the density distribution of the BEC is 
changed by the previous cycle. A cycle consisting of the excitation of the Rydberg 
atom followed by a finite interaction time with the condensate atoms is repeated 
$300$ or $500$ times as it was done in the experiment.  \\
\indent
The energy of the system increases by the time-dependent potential $V_{Ryd}$. Thus the 
condensate fraction is reduced. Some of the ground state atoms are promoted from the condensate 
to the thermal cloud. Within the CFA the two components of the 
bosonic gas -- the condensate and the thermal cloud are identified by accounting 
for the coarse graining as an unavoidable element of the measurement process \cite{CFA},
\footnote{A close analogy exists with the classical, Maxwell electrodynamics. At a microscopic level at each point in space and time an electric field has a well defined value even for the most complicated field. While a product of electric fields at two points in space-time is a well defined but usually useless number, a coarse graining caused by the detectors makes such a quantity useful, enabling, for instance, a proper definition of coherence length/time. See also a review article:  M. Brewczyk, M. Gajda, and K. Rz{\k a}\.zewski, J. Phys. B {\bf 40}, R1 (2007).}. 
Here, we define a coarse grained one-particle density matrix as resulting from 
the column integration \cite{Tomek}
\begin{equation}
\bar{\rho}(x,y,x',y';t) = \frac{1}{N}  \int dz \, \Psi(x,y,z,t) \, \Psi^*(x',y',z,t)  \,.
\label{average}
\end{equation}
where the $x$-axis is the condensate symmetry axis and the imaging is performed along
the radial direction. The resulting density matrix, upon spectral decomposition 
\cite{Penrose}, determines the fraction of the condensed atoms as a dominant 
eigenvalue. 

\begin{figure}[thb]
\includegraphics[width=8.cm]{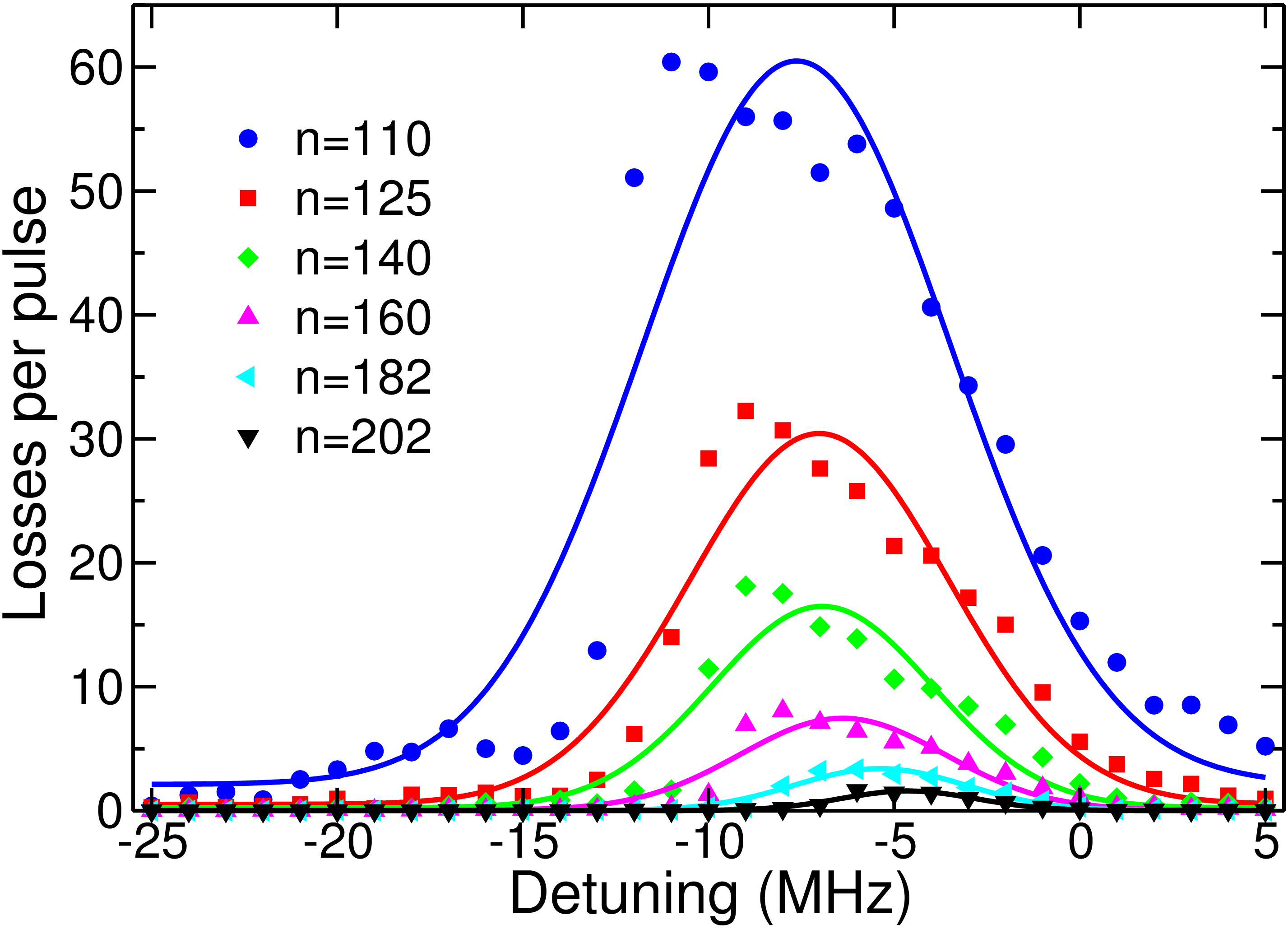}\\
\includegraphics[width=2.8cm]{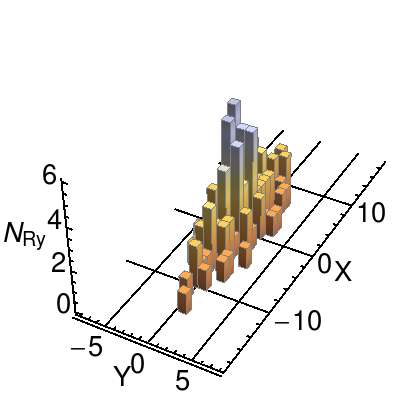}
\includegraphics[width=2.8cm]{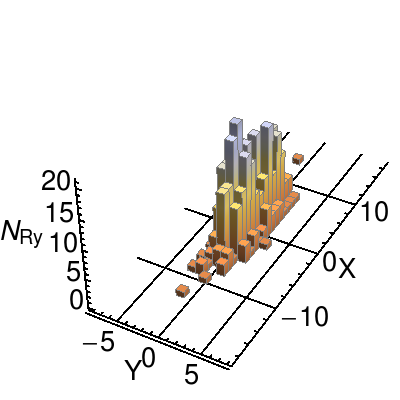}
\includegraphics[width=2.8cm]{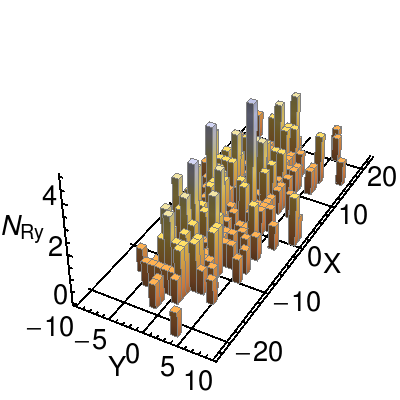}  \\ \vspace{0.3cm}
\includegraphics[width=2.8cm]{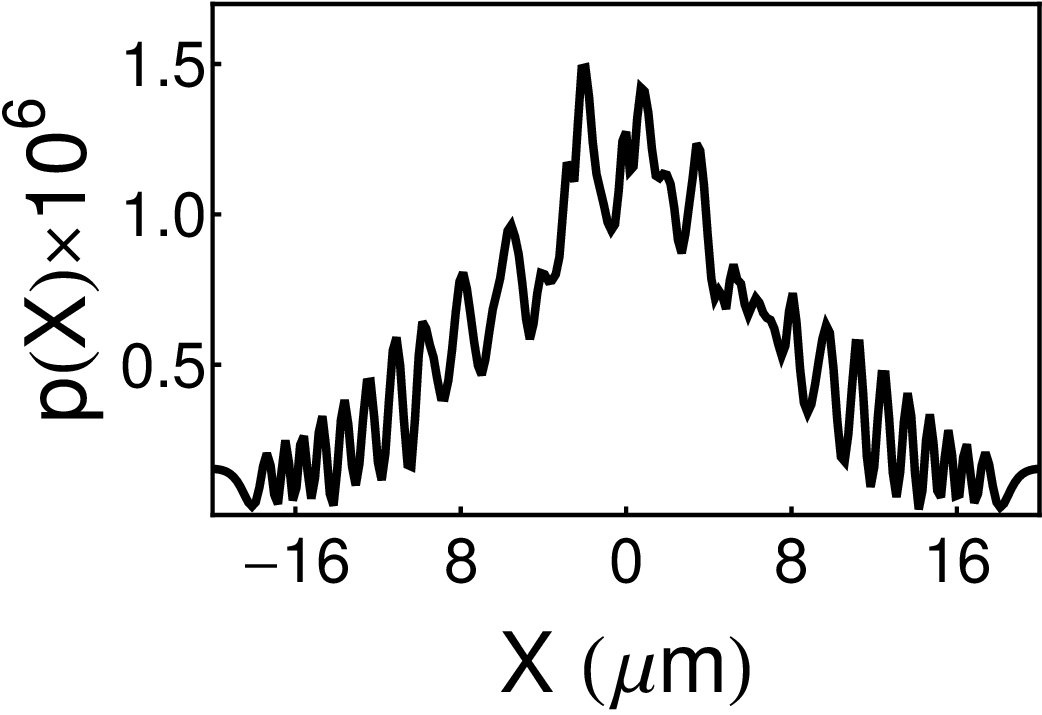}
\includegraphics[width=2.8cm]{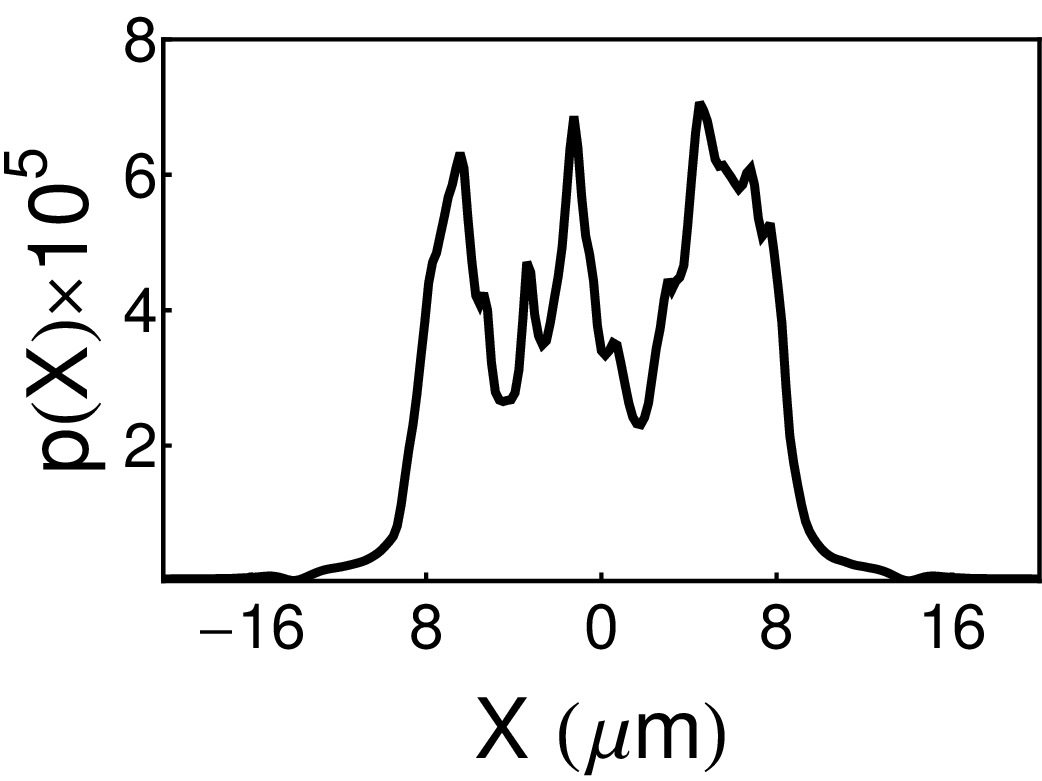}
\includegraphics[width=2.8cm]{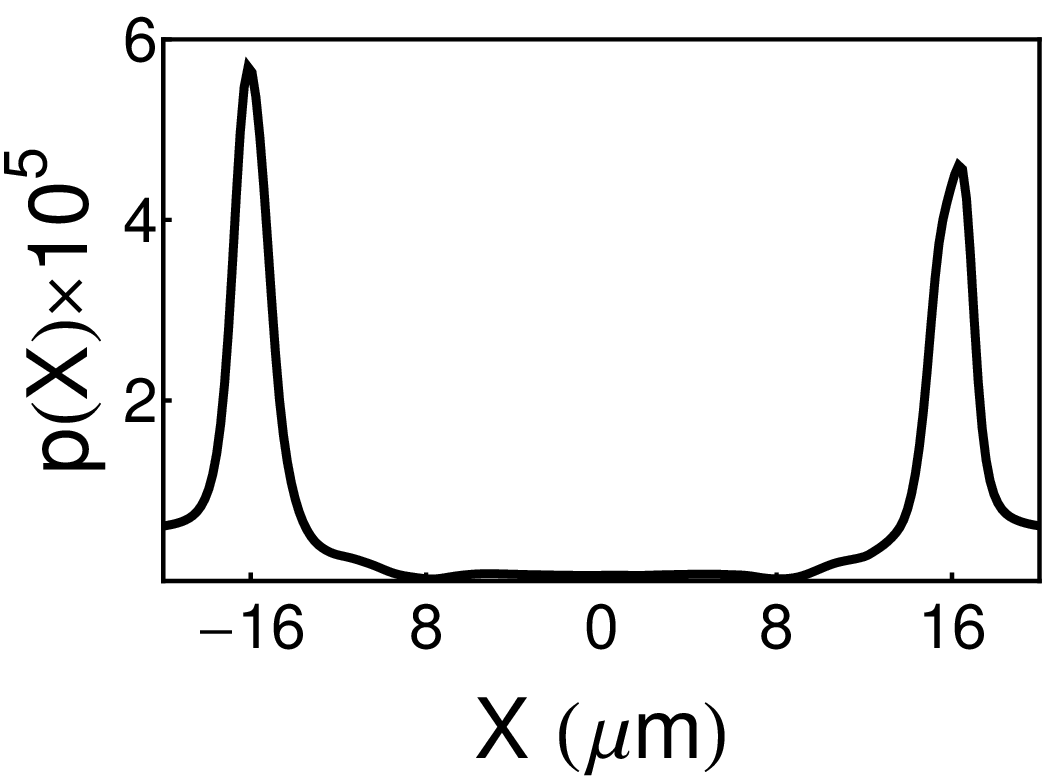}
\caption{(color online)
Theoretically modeled losses of atoms from the condensate per laser excitation pulse versus the detuning from the non-interacting \mbox{Rydberg} level (top frame). Solid lines are Gaussian fits.
Middle panel: A single realization real-space distribution of Rydberg atoms for $n=110$ and $\Delta \omega_{L} = -13\,$MHz (left frame), -9\,MHz (middle frame), and $1\,$MHz (right frame).
Bottom panel: Excitation probability, Eq. (\ref{probability}), along the trapping symmetry axis averaged over $500$ cycles (taken at the moments of time when Rydberg atoms are created). 
}
\label{LinesFig}
\end{figure}

\indent
We calculate the total condensate losses at the end of the excitation 
sequence, divide them by the number of excitation cycles and study the dependence 
of this quantity as a function of the laser detuning~$\Delta\omega_L$ and the 
principal quantum number~$n$ of the \mbox{Rydberg} state (Fig.~\ref{LinesFig}).
On the blue side of the resonance the Rydberg atom is created almost in every shot but losses are small because Rydberg atoms are excited in regions of low density, far from the center (see middle panel in Fig. \ref{LinesFig}, right frame). Towards the center of the line more Rydberg atoms are excited around the center of the trap where the density of the condensate is high. This leads to the increase of losses reaching a maximum approximately at the point where $\Delta \omega_{L}$ is equal to $\delta E (\vec{R},t) / \hbar$ calculated at the center of the trap. On the red side of the resonance still many Rydberg atoms are excited in the center of the trap (compare left and central frames of the middle panel in Fig. \ref{LinesFig}), however, not in every excitation cycle and thus the overall losses decrease. In the case of $n=110$ and $\Delta \omega_{L}=-12\,$MHz \textasciitilde two third of the excitation pulses creates a Rydberg excitation while only every seventh trial is successful at $\Delta \omega_{L}=-16\,$MHz. 
The asymmetry of the process with respect to the center of the line stems from the detuning which is a function of the local density.
As in \cite{LossesNature} our BEC exhibits quadrupole oscillations after the excitation sequence
is finished. However, the losses do not continue. 

\begin{figure}[thb]
\includegraphics[width=6.5cm]{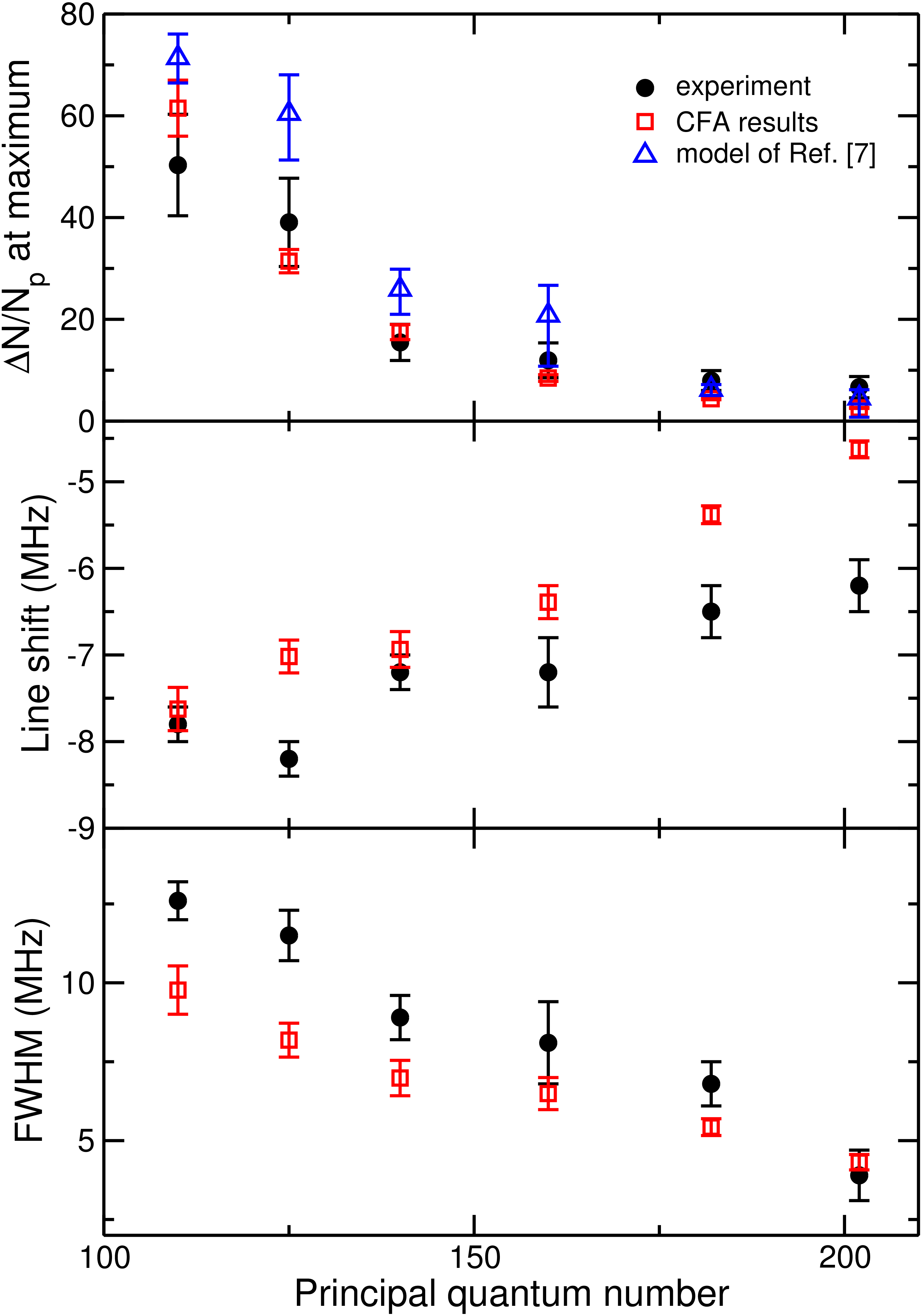}
\caption{(color online)
Comparison of theoretical results obtained within CFA (red open squares) and Bogoliubov approximation from Ref. \cite{LossesNature} (blue open triangles) with experimental data (black dots). The frames depict (from top to bottom) the maximum losses of atoms from the condensate per laser excitation pulse, the position of the resonance, and the FWHM of the resonance lines. The error bars of the CFA results are the statistical errors from the Gaussian fits.
}
\label{LDWFig}
\end{figure}

\begin{figure}[thb]
\includegraphics[width=7cm]{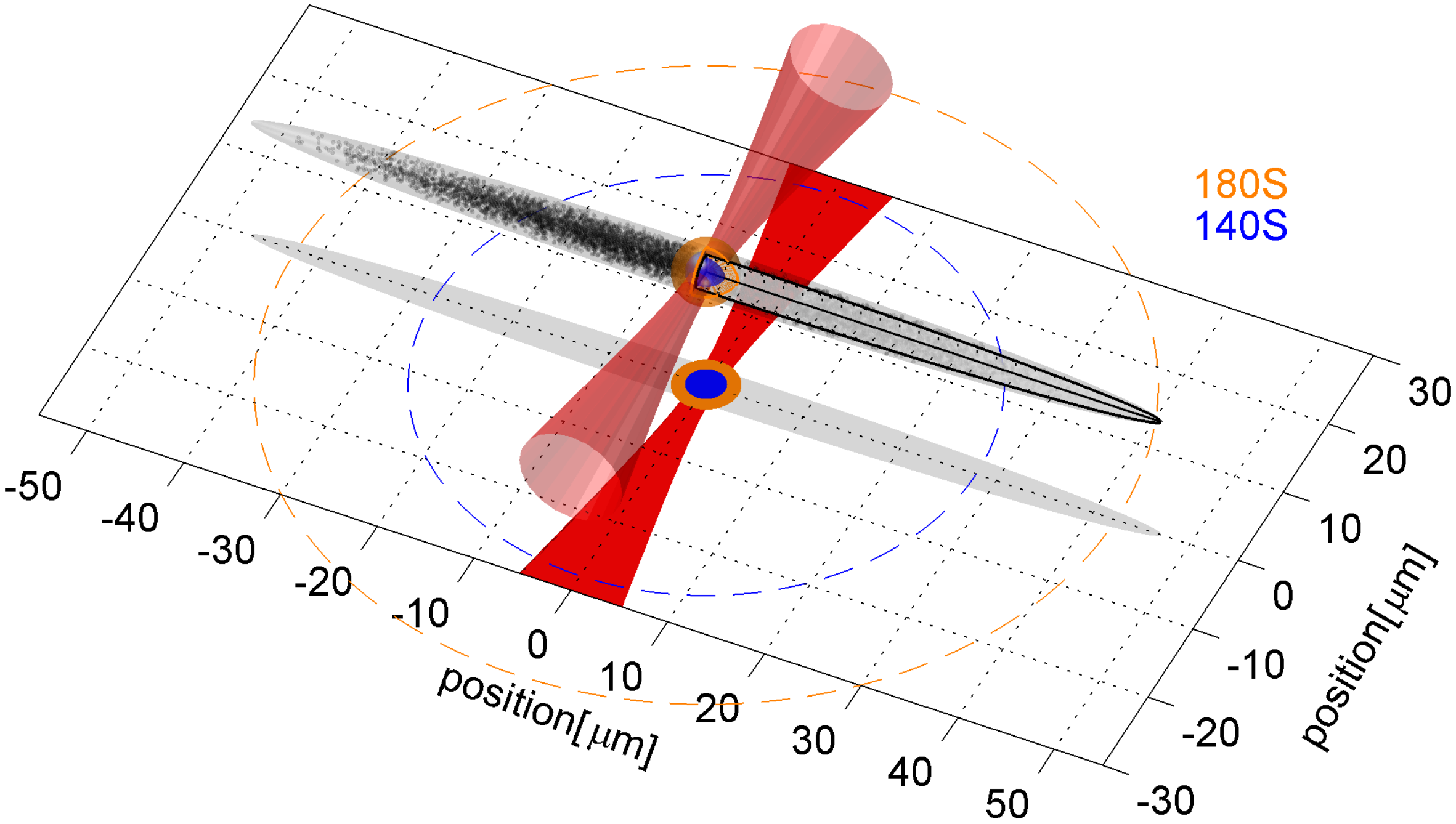}
\caption{(color online)
Rydberg atoms in 140S (blue) and 180D (orange) states are excited in the center of the condensate by a tightly focused laser beam (red). Dashed lines indicates the projection of the respective Rydberg blockade radii. All the sizes are to scale.
}
\label{Focus}
\end{figure}

\begin{figure}[thb]
\centering
\mbox{\includegraphics[width=4cm, trim=5 91 9 0, clip=true]{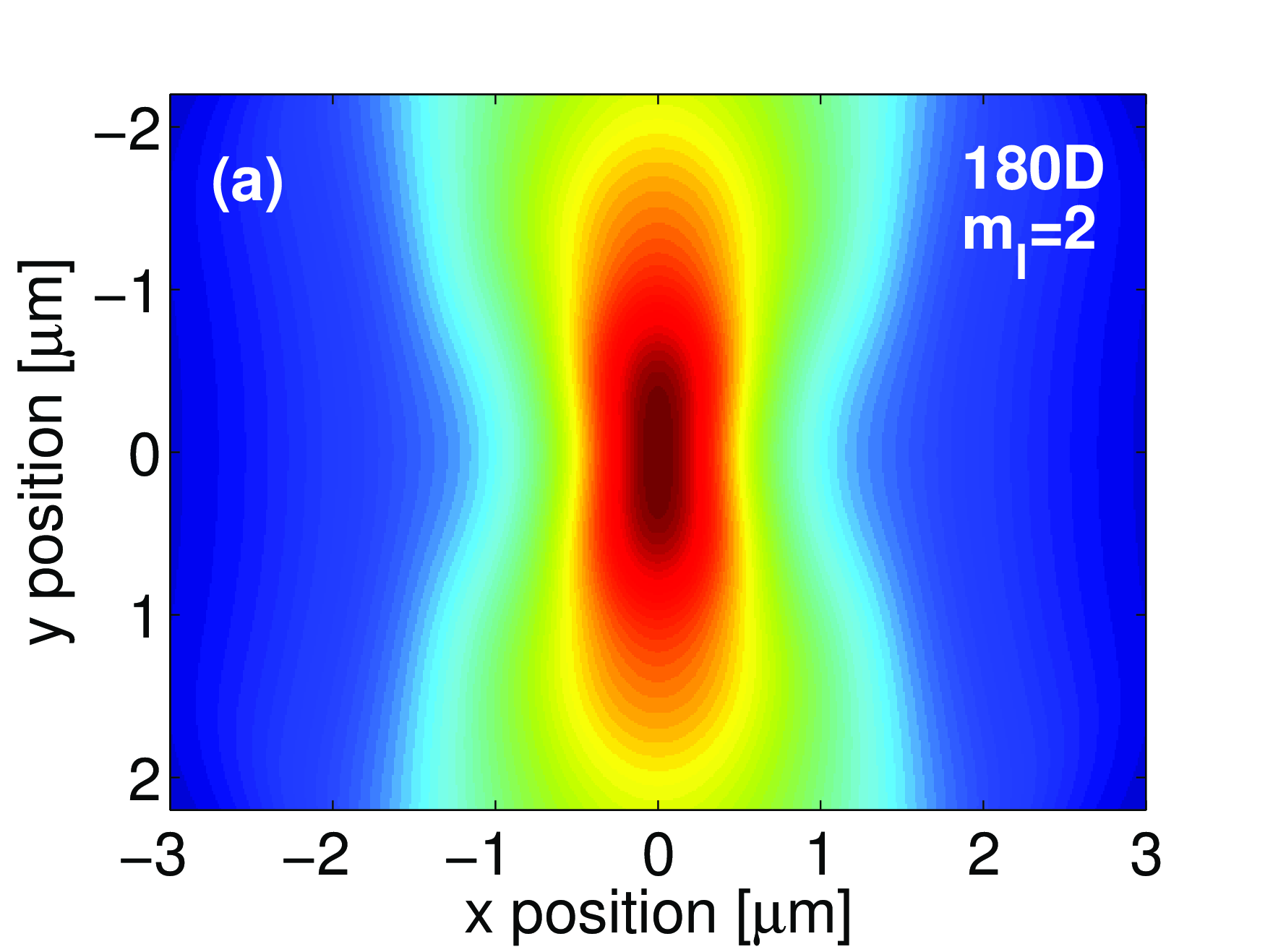} \hfill
\includegraphics[width=4cm, trim=5 91 9 0, clip=true]{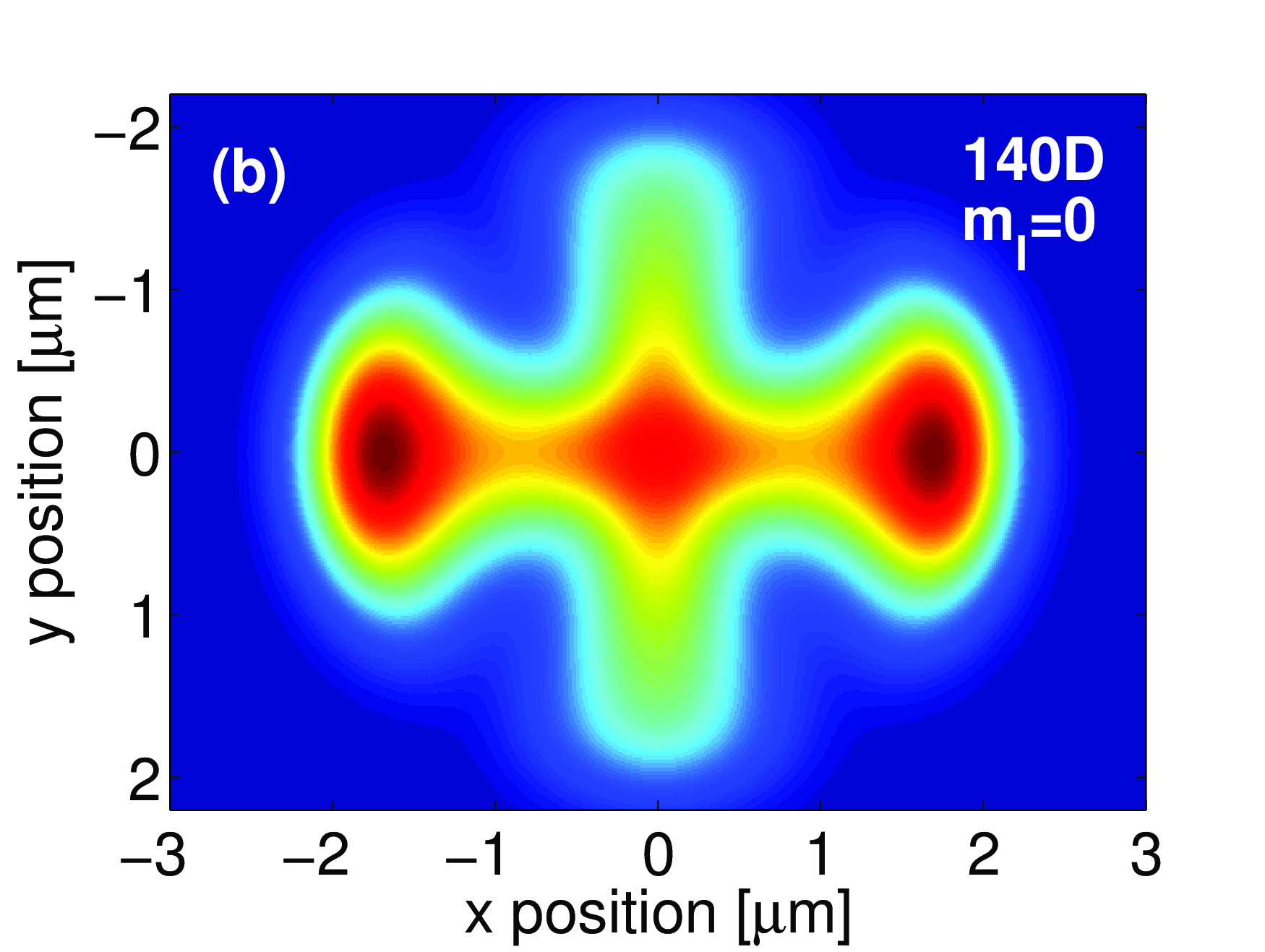}}  \\

\mbox{\includegraphics[width=4cm, trim=5 90 5 31, clip=true]{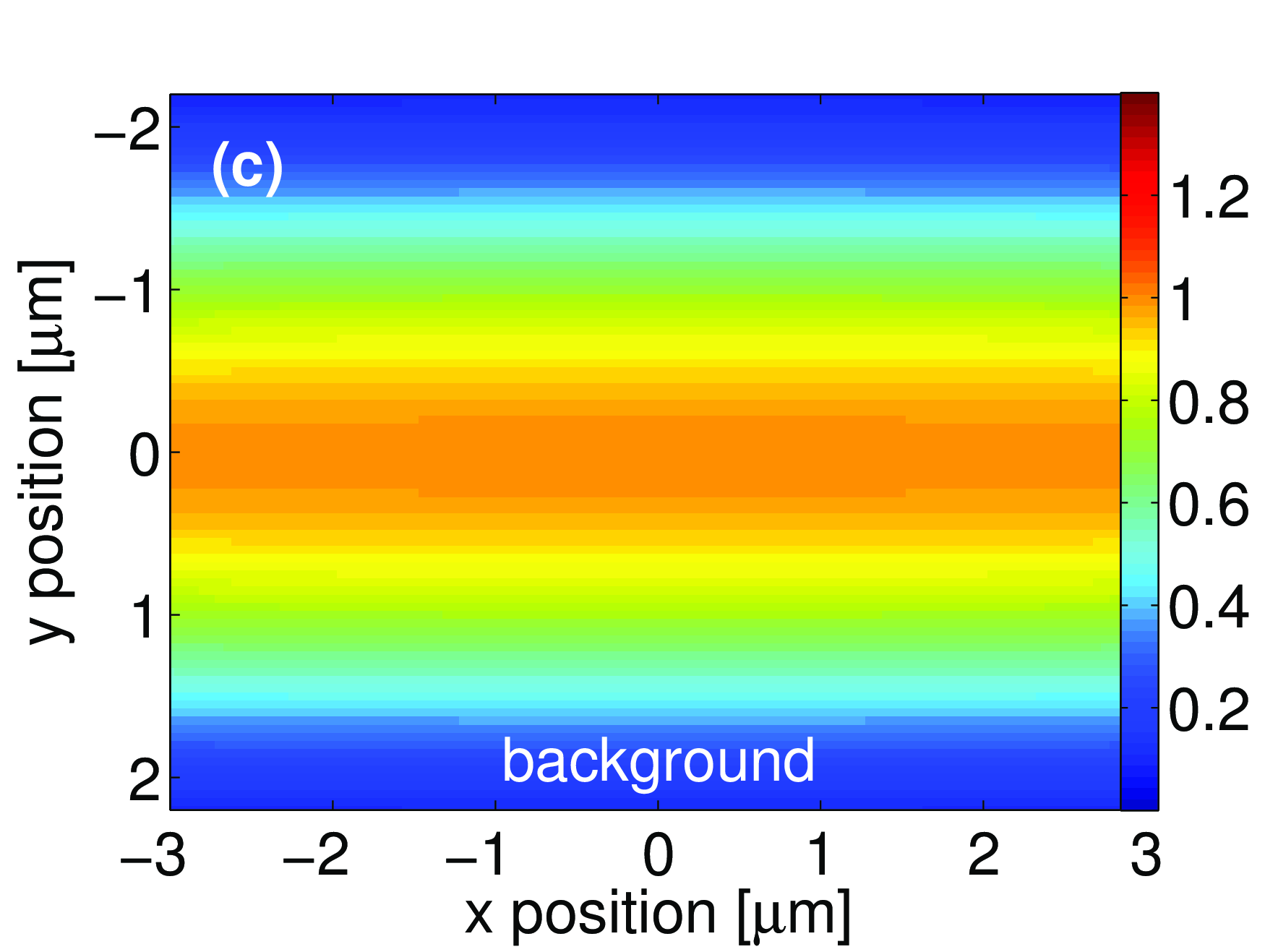} \hfill
\includegraphics[width=4cm, trim=5 90 5 31, clip=true]{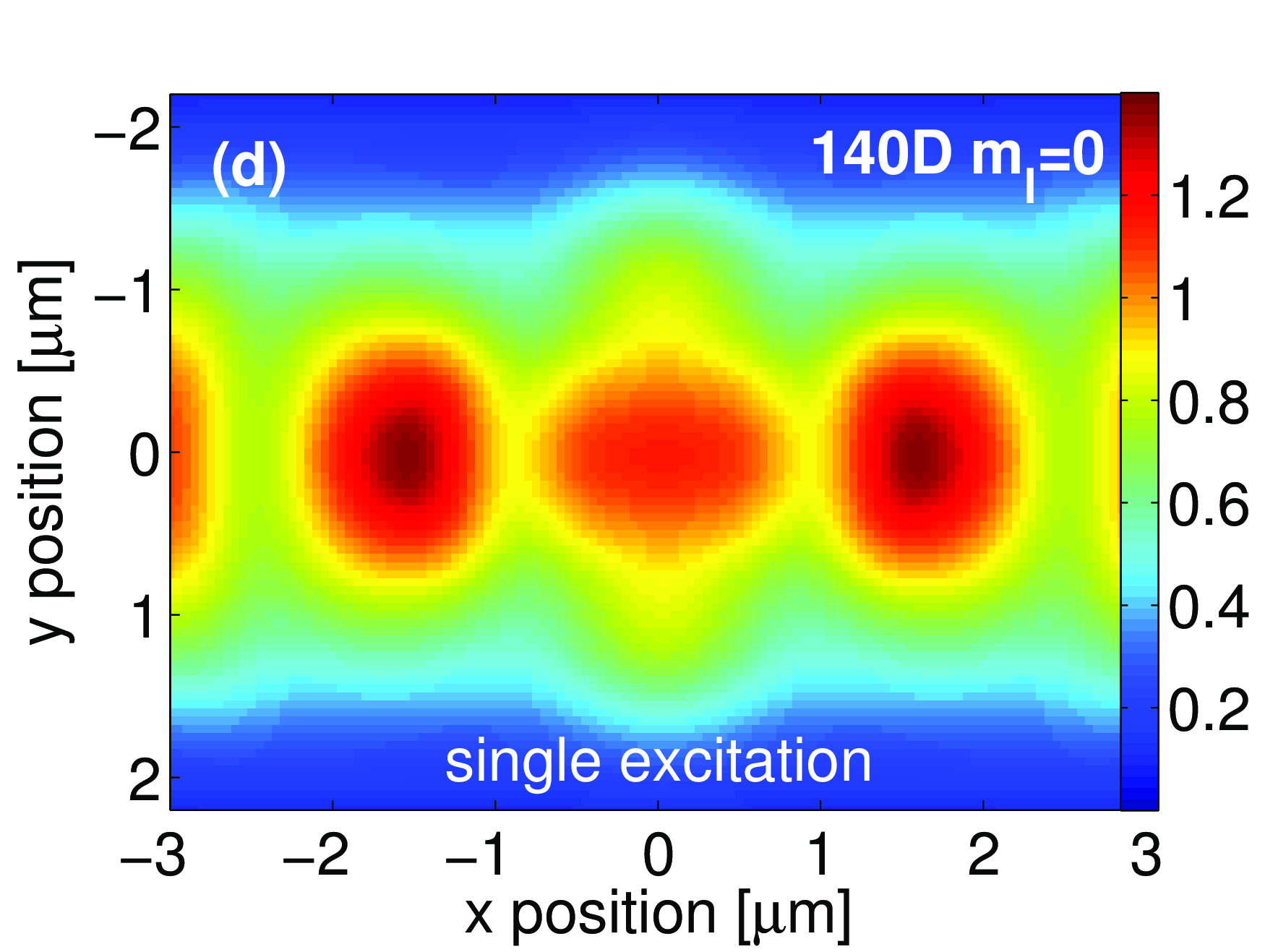}} 
\mbox{\includegraphics[width=4cm, trim=5 0 5 31, clip=true]{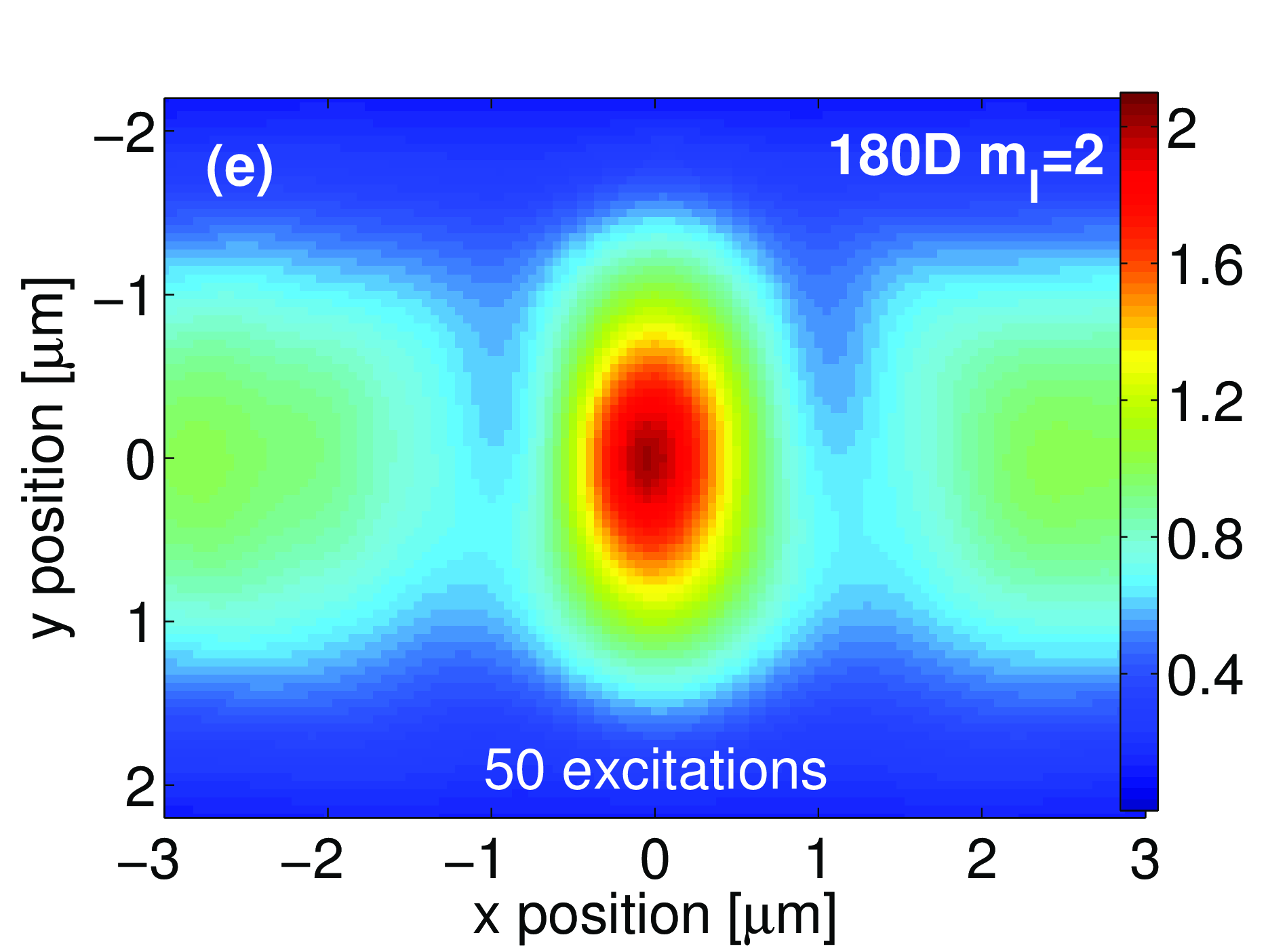} \hfill
\includegraphics[width=4cm, trim=5 0 5 31, clip=true]{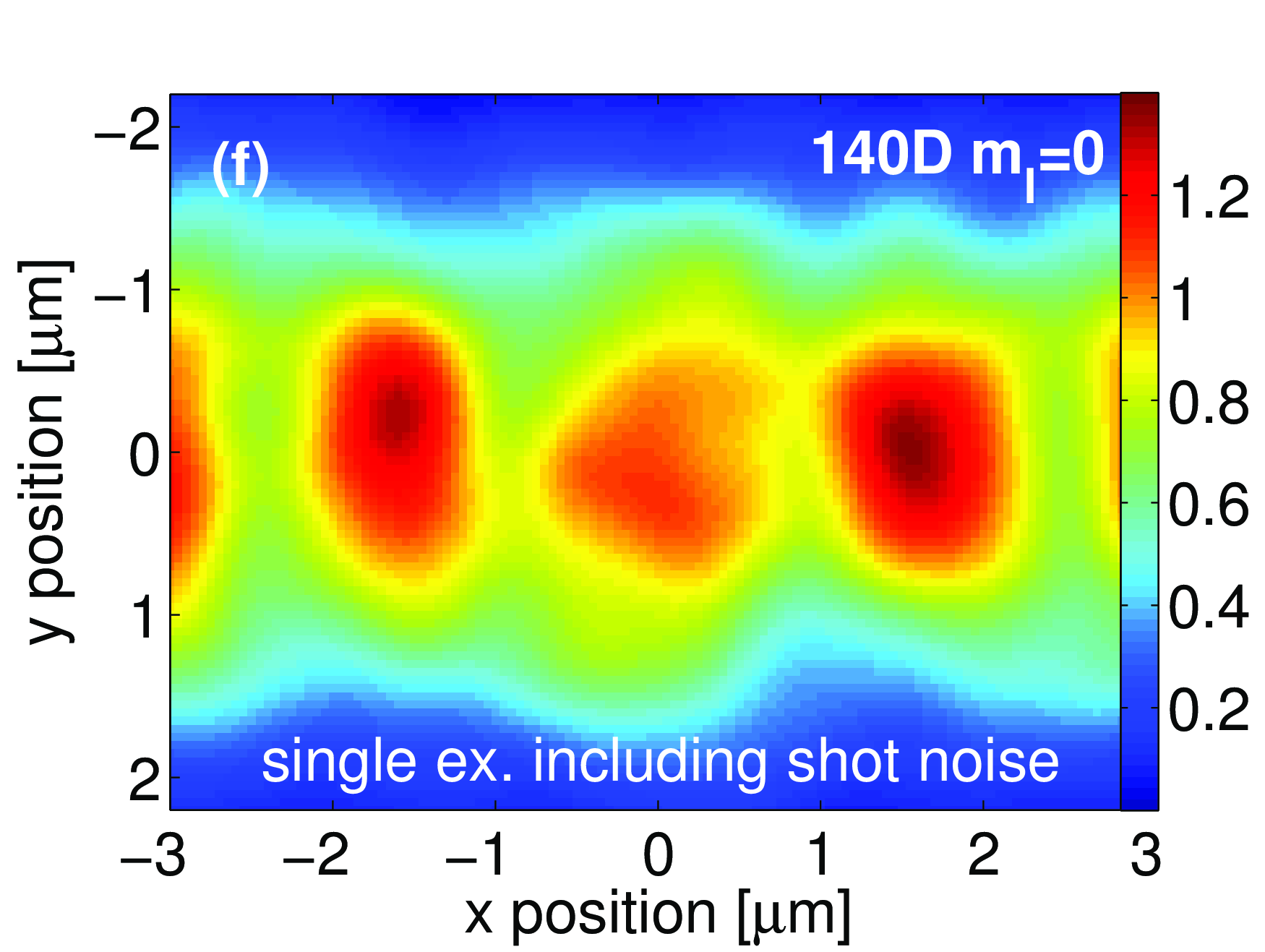}}
\caption{(a) and (b): Calculated orbitals for different Rydberg electrons convolved with a finite imaging resolution of $1\,{\mu}$m ($1/e^2$ width of the point spread function). This is the maximum contrast which can be expected from the imprint on the condensate. (c): center part of the BEC density distribution. The condensate consists of $5\cdot 10^4$ rubidium atoms and is confined in a harmonic trap with radial and axial frequencies $\omega_r=2\pi \cdot 200\,$Hz and $\omega_z=2\pi \cdot 10\,$Hz, respectively. (d) and (f): simulated density change caused by a single Rydberg atom without (d) and with atom number shot noise (f). These density patterns form after a single Rydberg atom lasting for $30\,{\mu}$s (which will be the case for every twentieth shot for a lifetime of $10\,{\mu}$s) and an additional evolution time of $180\,\mu$s. (e): density distribution of the BEC after $50$~Rydberg atoms have been consecutively excited in the center region ($1/e^2$ width: $1.2\,{\mu}$m).
}
\label{ImagingFig}
\end{figure}

\indent
The absolute values of maximal losses determined from the Gaussian fits to our 
numerical data (Fig.\ref{LinesFig}) are compared to experimental data and 
Bogoliubov calculations from \cite{LossesNature} in Fig.\ref{LDWFig} (top panel). 
We extract also the position of the resonance (middle panel) and the width of the 
line (bottom panel). The numerical results agree remarkably well with the experimental 
data considering the fact that only estimated values for the Rabi frequencies 
from the measurement and no additional free parameters were used.\\
\indent
While the overall atom loss was already quantitatively predicted within a Bogoliubov 
approach in \cite{LossesNature}, our method presented here provides additional 
insights and describes time evolution of the BEC during the experimental sequence 
in detail. This fact is of importance, since for every excitation but the first 
one, the condensate is already distorted due to the influence of the previous 
Rydberg atoms. Our method is nonperturbative and goes beyond first order approximation 
in phonon production. Moreover we predict the whole resonance line shape and 
include in the model the inhomogeneous density of the condensate caused by the 
trapping potential. 
\\
\indent
Having demonstrated that our model reproduces the experimental data very well,  
we now turn to the proposal of observing an electronic orbital by imaging 
the condensate density responding to the Rydberg potential~(eq. \ref{Ryd}). 
Our scheme relies on optical access with high numerical aperture as is readily 
available in many BEC experiments. Such high resolution optics enable the tight 
focusing of the excitation lasers into the center of the condensate, to define 
the position of the Rydberg atom(s) with high precision (Fig.\ref{Focus}). 
Moreover, this enables 
high resolution absorption images of the BEC to be taken. We consider Rydberg 
$s$- and $d$-states, which are accessible in typical two-photon excitation schemes 
\cite{Low, Viteau11}.\\
\indent
First, we study the case where the excitation lasers are kept on continuously 
(Fig.\ref{ImagingFig}e) to re-excite the Rydberg state as soon as the previous 
one has decayed. Here, the localization of the Rydberg atom must be within an area 
smaller than the expected structure size. Otherwise, the combined impact of many 
Rydberg excitations will wash out the Rydberg electron orbital imprint on the BEC. 
To resolve the overall structure of the exemplary 180D state (orbital radius $r=3.3\,\mu$m) 
it is sufficient to have the excitations within a diameter of $d=1.5\,\mu$m.
Sufficient sharpness and a good contrast of the image requires about 50 excitations, 
since the scattering potential depth is lower than the chemical potential of the 
condensate.
\\
\indent
To visualize the electron orbit of the Rydberg wavefunction with only one Rydberg 
excitation cycle (Fig.\ref{ImagingFig}d) the parameters of the experiment have to 
be chosen more carefully. A detailed study of this situation can be found in the~\cite{SM} 
and is summarized in the following. The principal quantum 
number of the Rydberg state must not be chosen too large because although the 
orbital radius scales with ${n^*}^2$, the effective potential drops with ${n^*}^{-6}$. 
The calculations show that the effective potential should be at least one order of 
magnitude deeper than the chemical potential of the atoms so that they can react 
during the lifetime of the Rydberg atom. This situation is reached for a principal 
quantum number around 140. Additionally, the thickness of the condensate, which the 
imaging light is traveling through, should not be larger than the orbital radius 
of the Rydberg electron. Otherwise the imaging light passes through an area that is
not affected by the imprint of the electronic wavefunction, which results in a 
reduction of contrast. In such a single-shot experiment the atom number shot noise 
\cite{AtomShotNoise} is the main source of noise. For the proposed parameter set 
(Fig.\ref{ImagingFig}f), a peak density $\rho\sim10^{14}\,$cm$^{-3}$ and a 
radial size of the condensate of $1.5\,\mu$m results in a $\sim6\,$\% background 
noise level (see supplementary material). This is well below the expected signal 
contrast of $\sim24\,\%$. 
Therefore, the Rydberg orbital imprint on the BEC density should be observable. 
Note, that the impact of the shot noise may be reduced by averaging images from 
multiple runs. 
\\
\indent
To summarize, we have presented and verified a theoretical, microscopic model of 
a single Rydberg electron in a Bose-Einstein condensate. Our theoretical model 
has several simplifications: 1. We do not account for the real losses of the 
trapped gas due to three body recombination. 2. The electron in its motion is not 
slow in the vicinity of the ionic core. We assume the electron-atom scattering 
length to be velocity independent, which  is a valid description beyond $n>80$. 
3. The impact of the ionic core on the heating process is neglected.\\
\indent
After verification of our theoretical model, we have proposed a novel scheme for 
mapping the electronic orbital onto the density of the condensate, thereby realizing 
a method to directly observe various electronic orbitals. 
Of course, with the available resolution we can image only the angular probability distribution
of a Rydberg orbital. Its radial structure will be washed out and this is not only due to
limited imaging resolution but also because tiny oscillating radial structure of higher Rydberg orbitals 
occurs on the space scale shorter than the healing length. 

\indent
Also exotic shapes of single electron probability densities in electric and magnetic fields 
including circular Rydberg states~\cite{HK83}, Stark states, Bohr-like wavepackets~\cite{MWL08, MGG09} 
and one dimensional atoms~\cite{HYB14} could be investigated in a way we propose. Furthermore, 
this approach can also be extended to more complex systems like Rydberg atom 
macrodimers~\cite{OST09} and multi-electron systems. Phase-sensitive images could 
be obtained if a structureless reference state is used in a coherent superposition 
state. The technical requirements with respect to resolution, both for the local 
excitation of \mbox{Rydberg} atoms and the detection of the resulting structures, 
are met by state of the art experimental setups. Furthermore, various techniques 
like dark ground imaging \cite{AMD96}, phase-contrast imaging~\cite{AKM97}, 
polarization contrast imaging~\cite{BSH97,KKS12} and adapted forms of absorption 
imaging~\cite{MOS12} are readily available to precisely determine the density 
distribution of a BEC in situ. The optical imaging of a single electron in a single 
shot experiment thus seems in direct reach.

\textsl{Acknowledgments:}
We are grateful to Mariusz Gajda and Tomasz Sowi\'{n}ski for helpful discussions. The work was supported by the National Science Center grants No. DEC-2011/01/B/ST2/05125 (T.K.) and DEC-2012/04/A/ST2/00090 (M.B., K.R.). K.R. acknowledges the financial support from the project ``Decoherence in long range interacting quantum systems and devices'' supported by contract research ``Internationale Spitzenforschung II'' of the Baden-W\"urttemberg Stiftung. The CQT is a Research Center of Excellence funded by the Ministry of Education and the National Research Foundation of Singapore. \\
The experimental work is funded by the Deutsche Forschungsgemeinschaft (DFG) within the SFB/TRR21 and the project PF 381/4-2. We also acknowledge support by the ERC under contract number 267100 and from E.U. Marie Curie program ITN-Coherence 265031. M.S. acknowledges support from the Carl Zeiss Foundation. S.H. is supported by the DFG through project HO 4787/1-1.

\bibliographystyle{apsrev4-1}
\bibliography{literature}

%\begin{thebibliography}{99}
%
%
%\bibitem{LossesNature} J.B. Balewski, A.T. Krupp, A. Gaj, D. Peter, H.P. B\"uchler, R. L\"ow, S. Hofferberth, and T. Pfau, Nature {\bf 502}, 664 (2013).
%
%\bibitem{Saffman} M. Saffman, T.G. Walker, K. M{\o}lmer, Rev. Mod. Phys. {\bf 82}, 2313 (2010).
%
%\bibitem{Fermi} E. Fermi, Nuovo Cim. {\bf 11}, 157 (1934).
%
%\bibitem{Amaldi} E. Amaldi and E. Segr\`e, Nature {\bf 133}, 141 (1934).
%
%\bibitem{Low} R. L\"ow, H. Weimer, J. Nipper, J.B. Balewski, B. Butscher, H.P. B\"uchler, and T. Pfau, J. Phys. B {\bf 45}, 113001 (2012).
%
%\bibitem{CFA} K. G\'oral, M. Gajda, and K. Rz\c a\.zewski, Opt. Express {\bf 10}, 92 (2001).
%
%\bibitem{Tomek} T. Karpiuk, M. Brewczyk, M. Gajda, and K. Rz\c a\.zewski, Phys. Rev. A {\bf 81}, 013629 (2010).
%
%\bibitem{Penrose} O. Penrose and L. Onsager, Phys. Rev. {\bf 104}, 576 (1956).
%
%
%
%\end{thebibliography}

\end{document}